\documentclass[12pt,a4paper]{article}
\newif\ifspringerclass\springerclassfalse

\usepackage{amsmath,amssymb,amsfonts}
\usepackage{graphicx}
\usepackage{booktabs}
\usepackage{array}
\usepackage{enumitem}
\usepackage{float}
\usepackage{xcolor}
\usepackage[hidelinks]{hyperref}

\ifspringerclass
\else
  \usepackage[a4paper,margin=1in]{geometry}
  \usepackage{authblk}
  \newcommand{\bmhead}[1]{\section*{#1}}
  \newcommand{\backmatter}{}
\fi

\hypersetup{ pdftitle={ECHO-PPI: Evidence-Bundled Discovery of Overlapping Protein Modules with Explainable Assignment Scoring}, pdfauthor={Sima Soltani, Reza Sheibani, Mehrdad Jalali, Yahya Forghani} }

\newcounter{algobox}
\newenvironment{algobox}[1]{%
  \refstepcounter{algobox}%
  \par\smallskip\noindent
  \begin{minipage}{\textwidth}
  \hrule
  \vspace{0.35em}
  \textbf{Algorithm~\thealgobox. #1}\par\smallskip
  \footnotesize
  \begin{enumerate}[leftmargin=1.6em,label=\arabic*.,itemsep=0.12em,topsep=0.12em]
}{%
  \end{enumerate}
  \vspace{0.2em}
  \hrule
  \end{minipage}
  \par
  \smallskip
}

\begin{document}

\begin{center}
{\LARGE\bfseries ECHO-PPI: Evidence-Bundled Overlapping Protein Module Detection with Hierarchical Assignment Confidence for Network Biology
\par}

\vspace{0.8em}

{\large Sima Soltani$^{1}$, Mehrdad Jalali$^{2,*}$, Yahya Forghani$^{1}$, Reza Sheibani$^{1}$\par}

\vspace{0.5em}

{\small
$^{1}$Department of Computer Engineering, Ma.C., Islamic Azad University, Mashhad, Iran\par
$^{2}$Applied Data Science and Artificial Intelligence, SRH University Heidelberg, Heidelberg, Germany\par
$^{*}$Corresponding author: Mehrdad.Jalali@srh.de\par
}
\end{center}

\vspace{1em}

\begin{abstract}
Identifying protein modules in protein--protein interaction (PPI) networks is central to understanding cellular organisation, yet many community-detection methods treat module membership as a binary output with limited assignment-level justification. Proteins that participate in multiple complexes as shared subunits, peripheral interactors, or context-dependent bridges can be overlooked when networks are forced into hard partitions, and even overlapping methods rarely provide traceable evidence for individual protein--module assignments. We present ECHO-PPI, a framework for overlapping protein-module detection that combines competitive module discovery with structured assignment-level interpretation.

For every protein--module assignment, ECHO-PPI exports an evidence bundle combining weighted topology, semantic functional similarity, Gene Ontology support, provenance fields, and hierarchical confidence labels: Core, Inner, Outer, and Uncertain. This makes each assignment inspectable and reproducible rather than an opaque membership claim. We benchmark ECHO-PPI on two yeast PPI resources, the Gavin socioaffinity network and the Krogan 2006 dataset, against MCL, MCL+overlap, ClusterONE, and SLPA. ECHO-PPI achieves predictive parity with overlap-aware baselines while being the only evaluated method to provide complete required-field evidence bundles. Core assignments show the strongest gold-standard support and consistent multi-channel evidence across both datasets. By separating predictive clustering from evidence-bundled interpretation, ECHO-PPI provides computational biologists with a path from cluster lists to defensible, reproducible protein-module hypotheses suitable for curator-facing network-biology workflows.
\end{abstract}

\noindent\textbf{Keywords:} Protein--protein interaction networks; Overlapping protein modules; Evidence-bundled discovery; Explainable assignment scoring; Gene Ontology; Interpretable bioinformatics

\bigskip
\section{Introduction}

High-confidence protein--protein interaction (PPI) networks summarize physical and functional associations inferred from affinity purification, yeast two-hybrid assays, and integrative databases such as STRING \cite{Szklarczyk2019}. A central computational-biology goal is to identify \emph{modules}---cohesive sets of proteins corresponding to molecular complexes, pathway components, or reusable functional units \cite{Gavin2006,Krogan2006}. Curated resources such as the EBI Complex Portal \cite{Meldal2022} enable benchmarking, but cellular organisation is not a hard partition: shared subunits and moonlighting proteins participate in multiple machineries \cite{Palla2005,Ahn2010}. The same protein can be a stable member of one complex, a peripheral participant in another, and a pathway-level bridge between processes. A useful PPI module-detection method therefore has to handle overlap without converting every possible overlap into an equally confident biological claim.

Benchmarking is also intrinsically incomplete. Gold-standard complexes are affected by database version, experimental ascertainment, identifier conventions, and the mapping between protein names used in PPI networks and names used in curated complexes. A predicted module can be biologically sensible yet fail a Jaccard threshold because the reference complex is partial; conversely, a module can match a gold standard while still containing boundary assignments that deserve scrutiny. For this reason, a single F1 score is not sufficient for curator-facing workflows. It should be accompanied by module-size statistics, coverage, overlap statistics, and, most importantly, a way to inspect why each protein--module assignment was made.

Classical and widely used PPI clustering methods can recover useful candidate modules, but they often separate detection from interpretation. Markov Clustering (MCL) \cite{vanDongen2000,Enright2002} remains widely used because it is fast and produces crisp modules on large graphs. Function-aware overlap heuristics can reassign boundary proteins using permanence and GO term statistics, yet they rarely attach \emph{per-protein} evidence explaining why a protein was placed in a module. Overlapping community methods---including clique percolation, link communities, and dense-subgraph approaches \cite{Adamcsek2006,Ahn2010,Nepusz2012,Bader2003}---allow multi-membership but seldom provide assignment-level audit trails suitable for curator review. As a result, biological interpretation is frequently performed after module detection, with limited provenance linking the final module list to the evidence that supported each assignment.

ECHO-PPI addresses the gap between \textbf{predictive clustering} and \textbf{auditable assignment}. The framework combines (i)~topology from weighted PPI graphs, (ii)~semantic embeddings of GO and text profiles \cite{Reimers2019,Ashburner2000,GeneOntology2021}, (iii)~evidence-potential nucleus scoring inspired by gravity-based representative selection \cite{Jalali2025BlackHole}, (iv)~broad candidate generation and multi-evidence scoring, (v)~overlap-aware seeding, (vi)~recall-safe supplementation, and (vii)~hierarchical confidence labels with evidence bundles. We report benchmarks conservatively: on Gavin, ECHO-PPI remains near MCL and MCL+overlap but does not match exact ClusterONE on predictive F1. The contribution is a reproducible audit layer for overlapping assignments, not a claim that every exported overlap is experimentally validated.

The main contributions are: (i) an evidence-bundled framework for overlapping PPI module discovery; (ii) integration of network-topological, functional, semantic, and provenance information at assignment and module level; (iii) reproducible module-level reporting for biological interpretation and curator triage; and (iv) benchmarking against established PPI and network module-detection baselines, including MCL, MCL+overlap, ClusterONE, SLPA, and related overlapping-community families. In this manuscript, ``trustworthy'' is used only in this restricted technical sense: evidence-constrained, reproducible, and inspectable computational decision support.

\section{Literature review}

Early high-throughput PPI studies showed both the promise and noise sensitivity of interaction maps \cite{vonMering2002}. In yeast, Gavin \textit{et al.}\ \cite{Gavin2006} introduced socioaffinity scores from tandem affinity purification, and Krogan \textit{et al.}\ \cite{Krogan2006} provided a complementary global complex map. Large interaction repositories such as BioGRID \cite{Stark2006} and STRING \cite{Szklarczyk2019} extend coverage by integrating heterogeneous experimental and computational evidence. Benchmarking then depends on curated complex resources, including yeast catalogues such as CYC2008 \cite{Pu2009CYC2008}, mammalian resources such as CORUM \cite{Giurgiu2019}, and the EBI Complex Portal \cite{Meldal2022}. ECHO-PPI follows common complex-matching practice by evaluating predicted modules against curated references with greedy Jaccard matching at threshold $0.5$ \cite{Brohee2006,Pu2019}. Recent review and benchmark studies reinforce the need for evaluation beyond a single score: Patra and Sahoo summarise persistent PPI-complex challenges including noise, sparse and small complexes, overlap, scalability, and biological context \cite{patra2026review}, while PRING argues that PPI predictors should be assessed at graph level because pairwise accuracy alone may not preserve topology, functional modules, or downstream utility \cite{zheng2025pring}.

Hard graph clustering remains a strong baseline. MCL simulates stochastic flow on a powered adjacency matrix, with inflation controlling granularity \cite{vanDongen2000,Enright2002}; it is fast and reproducible but returns a hard partition. Modularity optimisation methods such as Louvain \cite{Blondel2008} and Leiden \cite{Traag2019} scale well and expose multiresolution structure, but they are also primarily partition-oriented and can produce communities whose biological interpretation depends strongly on resolution and preprocessing choices. These methods are useful baselines for PPI maps, yet they rarely explain why a particular protein belongs to a particular module.

Overlapping community detection better matches cellular biology because shared subunits can participate in multiple complexes. Clique percolation and CFinder identify adjacent dense cliques \cite{Palla2005,Adamcsek2006}; link communities assign edges to modules and induce node overlap \cite{Ahn2010}; ClusterONE grows dense PPI subgraphs with overlap support \cite{Nepusz2012}; OSLOM searches for statistically significant local clusters \cite{Lancichinetti2009}; label-propagation variants such as SLPA produce soft or overlapping assignments \cite{Xie2011SLPA}; and overlapping modularity formulations quantify overlap quality \cite{Nicosia2009}. Recent biological-network work continues this direction: RDS combines diffusion-state-distance spectral clustering with ReCIPE post-processing to reconnect disconnected biological clusters and introduce limited overlap on yeast and human protein association networks \cite{ocitti2025rds}. General surveys emphasise the scalability of label-propagation approaches but also their sensitivity to propagation choices and evaluation design \cite{goswami2025labelPropagation}; broader network-science methods such as CD-DAWN extend overlap detection to directed weighted graphs, but are not direct baselines for the undirected weighted yeast PPI benchmarks used here \cite{kumar2025cddawn}. These approaches make multi-membership possible, but their outputs are usually module lists rather than assignment-level evidence records.

Functional evidence can improve biological plausibility, but it must be handled carefully. GO provides structured annotations for molecular function, biological process, and cellular component \cite{Ashburner2000,GeneOntology2021}; PPI-based function prediction has long combined graph topology with indirect-neighbour or weighted-neighbour evidence \cite{Chua2006}. Recent GO-aware optimisation work similarly indicates that biological annotations can improve complex detection when integrated into the search process, for example through GO-based mutation operators in evolutionary algorithms \cite{abbas2025goMutation}. Modern text embeddings such as Sentence-BERT provide dense semantic profiles \cite{Reimers2019}. Graph-embedding approaches have also entered overlapping protein-complex detection: GAER-GMM combines graph autoencoder representations, Gaussian mixture modelling, and protein-related biological features to identify overlapping complexes \cite{tu2025gaergmm}. ECHO-PPI uses GO and semantic similarity differently: as explicit support channels exported for each protein--module assignment, while retaining graph gates so that annotation similarity alone does not silently promote disconnected proteins into confident complex membership.

Several PPI methods include local quality ideas, such as vertex permanence for community-membership strength \cite{Chakraborty2014Permanence}, but few export complete per-assignment records for manual quality control. The Black Hole Strategy \cite{Jalali2025BlackHole} ranks representative nodes in graphs using gravity-inspired scores; ECHO-PPI adapts this idea as a \emph{deterministic} evidence-potential nucleus score for candidate generation, not as a physical simulation. The framework is therefore positioned between predictive clustering and auditable curation: it keeps the strong overlap-seeded baseline but adds confidence labels and evidence bundles for every protein--module assignment.

Table~\ref{tab:positioning} contrasts ECHO-PPI with common approaches. The novelty is not universal F1 superiority, but evidence-constrained and auditable overlapping assignment with hierarchical confidence labels.

\begin{table}[t]
\centering
\caption{Positioning of ECHO-PPI relative to prior PPI community-detection approaches.}
\label{tab:positioning}
\scriptsize
\resizebox{\textwidth}{!}{%
\begin{tabular}{p{2.4cm}cccccc}
\toprule
Approach & Overlap & PPI use & Bio. evidence & Module interpretation & Provenance & Ranked bundles \\
\midrule
MCL & No & Common & No & Post-hoc & Limited & No \\
MCL + overlap & Limited & Yes & GO TF--IDF & Partial & Limited & No \\
ClusterONE & Yes & Designed for PPI complexes & Post-hoc & Module list & Limited & No \\
CFinder / clique percolation & Yes & Used in biological networks & Rarely & Clique modules & Limited & No \\
Link communities & Yes & Used in PPI networks & Rarely & Edge-derived modules & Limited & No \\
Louvain / Leiden & No & Common baseline & No & Partition modules & Limited & No \\
SLPA & Yes & General overlapping baseline & Rarely & Label memberships & Limited & No \\
RDS / ReCIPE & Yes & Biological association networks & Functional enrichment / post-hoc & Reconnected clusters & Partial & No \\
GAER-GMM & Yes & PPI complex detection & Biological features & Predicted complexes & Partial & No \\
\textbf{ECHO-PPI} & \textbf{Yes} & \textbf{PPI-specific} & \textbf{Topology, GO, semantic evidence} & \textbf{Labels and evidence summaries} & \textbf{Assignment-level fields} & \textbf{Yes} \\
\bottomrule
\end{tabular}
}%
\end{table}

\section{Methods}

\subsection{Problem formulation}
Let $G=(V,E,w)$ be an undirected weighted PPI graph, where $V$ is the protein set, $E$ is the interaction set, and $w:E\to\mathbb{R}_{\ge 0}$ maps each interaction to a confidence or socioaffinity weight. A community-detection method returns modules $\mathcal{C}=\{C_1,\ldots,C_m\}$, where $C_i\subseteq V$ and $|C_i|\geq 2$. Unlike hard clustering, ECHO-PPI allows $\sum_i \mathbb{I}(p\in C_i)>1$ for proteins with plausible multi-membership.

Reference complexes $\mathcal{G}=\{G_k\}$ define evaluation. Predicted modules are matched to reference complexes using the Jaccard index
\begin{equation}
\label{eq:jaccard}
\mathrm{J}(C_i,G_k)=\frac{|C_i\cap G_k|}{|C_i\cup G_k|}.
\end{equation}
Greedy maximum-weight matching sorts all predicted--gold pairs by $\mathrm{J}$, accepts an unmatched pair if $\mathrm{J}\geq 0.5$, and then marks both modules as used. Precision is the fraction of predicted modules that are matched, recall is the fraction of gold complexes that are matched, and F1 is their harmonic mean. This protocol is common in PPI complex benchmarks \cite{Brohee2006,Pu2019,Meldal2022}; it is strict enough to penalise diffuse expansions, but it should not be interpreted as a complete measure of biological validity.

\subsection{Data and preprocessing}
We benchmark on the Gavin yeast socioaffinity network after removing spreadsheet-corrupted identifiers such as \texttt{\#NAME?} and node labels with embedded whitespace. The cleaned primary graph contains 1,848 proteins and 7,578 weighted interactions. Table~\ref{tab:dataset-availability} summarises the benchmark sizes, GO coverage, gold-standard coverage, and mapping status used in the reported evaluation. Edge weights are min--max normalised to $[0,1]$. Generic GO roots (molecular function, cellular component, biological process) are excluded from scoring. Reference complexes are mapped to locus identifiers; complexes with fewer than two proteins are excluded from matching.

As a second fully reproducible benchmark, we extracted the Krogan \textit{et al.}\ Nature 2006 yeast interaction set \cite{Krogan2006} from the local BioGRID TAB3 file (accessed 2026-05-18) by filtering records to \texttt{PUBMED:16554755}, physical yeast--yeast interactions, and SGD systematic ORF identifiers. The preprocessing removes self-loops, deduplicates undirected edges, preserves numeric BioGRID scores when present and otherwise assigns unit edge weight. This produces a Krogan graph with 2,671 proteins and 7,072 interactions. Dataset-specific cleaning, identifier-mapping, and GO-coverage reports are exported under \texttt{results/krogan/}.

GO annotation coverage is computed after harmonising SGD records to the systematic yeast ORF identifiers used by the Gavin graph. The parser indexes DB object identifiers, gene symbols, and ORF-like synonyms in the GAF synonym field, then removes only the three generic GO roots from scoring. Under this mapping, 1,834 of 1,848 cleaned Gavin nodes retain at least one non-generic GO annotation. Residual missing GO evidence is treated as missing evidence rather than as biological absence, and topology and semantic support are reported separately from GO support.

The codebase also contains loaders or feasibility hooks for STRING and broader BioGRID yeast networks. In the present manuscript, Gavin remains the primary benchmark and Krogan is used as a second transferability benchmark. Secondary dataset loader and mapping feasibility notes for STRING and the full BioGRID network are reported in Appendix~\ref{app:secondary-feasibility}.

\begin{table}[t]
\centering
\caption{Primary benchmark and identifier-mapping status used for the reported ECHO-PPI evaluation.}
\label{tab:dataset-availability}
\scriptsize
\begin{tabular}{lrrrrp{3.7cm}}
\toprule
Dataset & Edges & Nodes & GO nodes & Gold cov. & Status \\
\midrule
Gavin yeast & 7,578 & 1,848 & 1,834 & 0.579 & Cleaned primary benchmark \\
Krogan yeast & 7,072 & 2,671 & 2,652 & 0.699 & Krogan 2006 transferability benchmark \\
\bottomrule
\end{tabular}
\end{table}

\subsection{ECHO-PPI overview}
\begin{figure}[htbp]
\centering
\IfFileExists{Figure_1.png}{%
\includegraphics[width=0.95\textwidth]{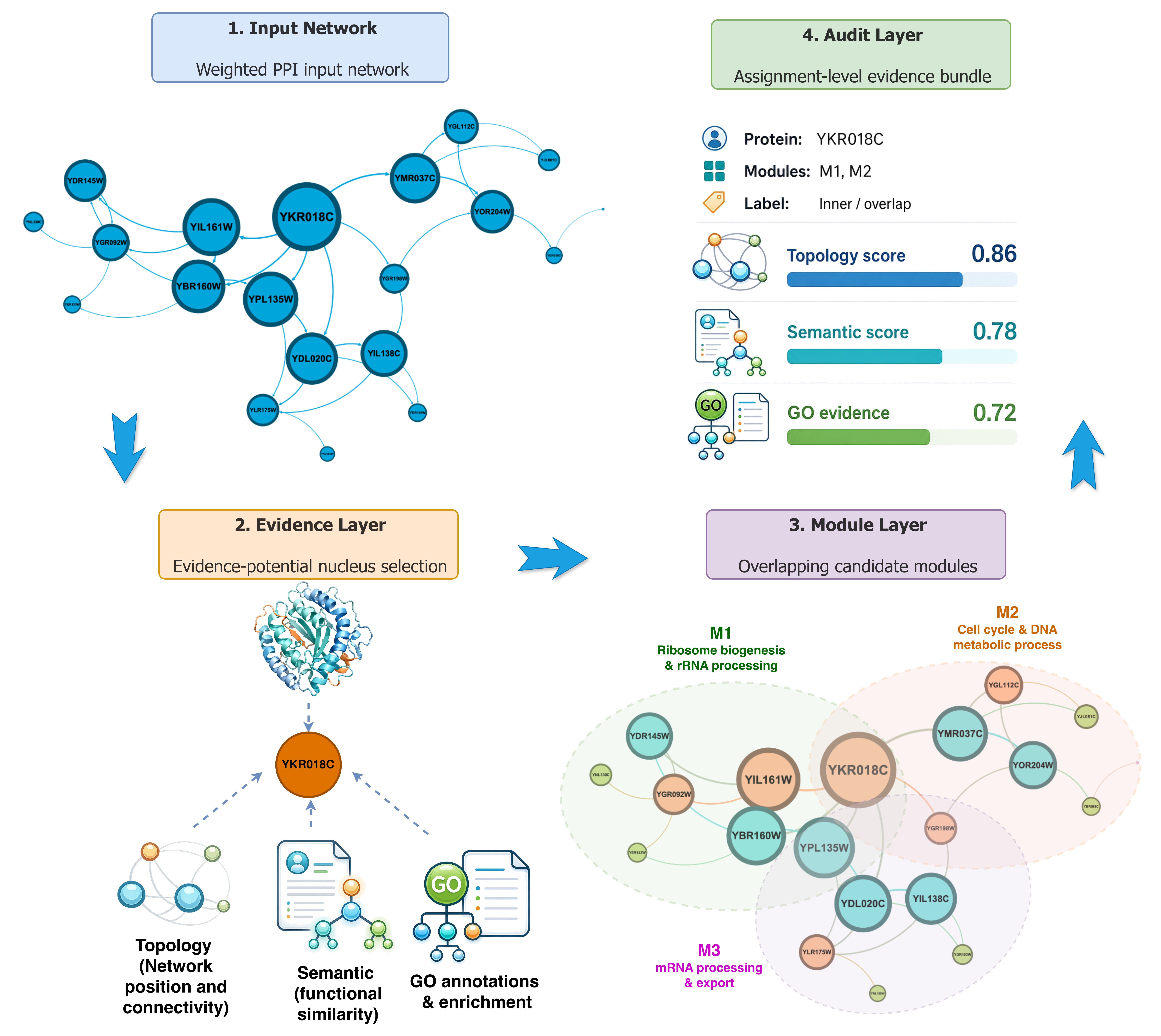}%
}{%
\IfFileExists{Figure_1.png}{%
\includegraphics[width=0.95\textwidth]{Figure_1.png}%
}{%
\fbox{\parbox{0.9\textwidth}{\centering Figure 1 file not found. Keep the original author-approved figure and place it as \texttt{Figure\_1.png}.}}%
}%
}
\caption{ECHO-PPI four-layer workflow. A weighted PPI input network is transformed into evidence-potential nucleus selection using topology, semantic functional similarity, and GO annotation/enrichment signals. The module layer constructs overlapping candidate modules, with YKR018C shown as a simplified illustrative M1--M2 overlap example. The audit layer exports assignment-level evidence bundles with confidence labels and channel-specific evidence scores.}
\label{fig:workflow}
\end{figure}

Figure~\ref{fig:workflow} illustrates the four-layer ECHO-PPI workflow. The input layer starts from a weighted PPI network, where edge widths represent interaction confidence. The evidence layer integrates topology, semantic functional similarity, and GO annotation/enrichment signals to select an evidence-potential nucleus. The module layer constructs overlapping candidate modules and highlights YKR018C as a simplified illustrative protein assigned to both M1 and M2. The audit layer exports the corresponding assignment-level evidence bundle, including the protein identifier, assigned modules, confidence label, and channel-specific evidence scores.

\subsection{Technical novelty}
ECHO-PPI is not presented as a generic application of artificial intelligence to PPI clustering. Its technical contribution is the coupling of overlap-aware module discovery with structured, reproducible evidence reporting. First, the method keeps protein multi-membership explicit by constructing and scoring overlapping candidate modules rather than forcing a single partition. Second, every retained protein--module assignment is exported with topology, semantic, GO, confidence-label, and provenance fields, so that interpretation remains connected to the computational decision that produced the assignment. Third, the evidence-potential nucleus score provides a deterministic ranking mechanism for local candidate generation while leaving the final module decision to overlap-aware scoring and conservative supplementation. Fourth, module-level reports are designed to be reproducible: dataset cleaning, identifier harmonisation, evidence-channel scores, matching thresholds, and confidence-label rules are all recorded rather than implied.

This design targets a specific computational-biology use case: exploratory PPI module analysis in which predicted overlapping modules require biological interpretation, filtering, and audit. The novelty is therefore traceable interpretation of protein-module assignments, not a claim that an evidence bundle itself proves physical complex membership.

\subsection{Evidence profiles and embeddings}
For each protein $p\in V$ we compute normalised weighted degree, clustering coefficient, $k$-core score, non-generic GO richness, and annotation-uncertainty proxies. Text profiles concatenate locus identifiers and GO labels. All reported benchmarks used the TF--IDF+SVD backend with 64 dimensions; a Sentence-BERT backend is available when \texttt{sentence-transformers} and the model cache are installed, but it was not used for the reported results.

\begin{algobox}{Evidence profile construction}
\label{alg:profiles}
\item For each protein $p\in V$, compute topology features from $G$.
\item Normalise weighted degree, clustering coefficient, and $k$-core score.
\item Count non-generic GO terms and compute an annotation-uncertainty proxy.
\item Build the text profile and store the resulting multimodal feature vector.
\end{algobox}

\subsection{GO TF--IDF functional signatures}
ECHO-PPI treats modules as documents and GO terms as words when constructing functional signatures. For a GO term $t$ in module $C_i$, term frequency is
\begin{equation}
\label{eq:tf}
\mathrm{TF}(t,C_i)=\frac{|\{p\in C_i:t\in G(p)\}|}{|C_i|},
\end{equation}
where $G(p)$ is the non-generic GO term set for protein $p$. Inverse document frequency downweights generic terms that appear in many modules:
\begin{equation}
\label{eq:idf}
\mathrm{IDF}(t)=\log\frac{|\mathcal{C}|}{|\{C_i\in\mathcal{C}: \exists p\in C_i,\ t\in G(p)\}|}.
\end{equation}
The cluster-specific functional signature is then
\begin{equation}
\label{eq:tfidf}
\mathrm{TFIDF}(t,C_i)=\mathrm{TF}(t,C_i)\times \mathrm{IDF}(t).
\end{equation}
High TF--IDF terms are frequent within a module but rare across other modules, making them useful for functional dependency scoring and evidence summaries. Missing annotations are treated as missing evidence, not evidence of biological absence.

\begin{algobox}{GO TF--IDF calculation}
\label{alg:tfidf}
\item Input: modules $\mathcal{C}$ and protein--GO map $G$.
\item For each module $C_i\in\mathcal{C}$, collect all non-generic GO terms assigned to proteins in $C_i$.
\item Increment $\mathrm{count}(t,C_i)$ and mark term $t$ as present in document $C_i$.
\item For each term--module pair, compute $\mathrm{TF}(t,C_i)$, $\mathrm{IDF}(t)$, and $\mathrm{TFIDF}(t,C_i)$ using Eqs.~\eqref{eq:tf}--\eqref{eq:tfidf}.
\item Return the module--term TF--IDF matrix.
\end{algobox}

\subsection{Permanence and boundary topology}
Permanence quantifies how well protein $p$ is topologically embedded in module $C_i$ relative to other modules. Let $N(p)$ be the neighbours of $p$, $I(p,C_i)=|N(p)\cap C_i|$ be its internal degree with respect to $C_i$, and
\begin{equation}
\label{eq:emax}
E_{\max}(p,C_i)=\max_{C_j\neq C_i}|N(p)\cap C_j|
\end{equation}
be its maximum external connectivity to any competing module. Let $C_{\mathrm{in}}(p,C_i)$ be the clustering coefficient among the internal neighbours $N(p)\cap C_i$. The permanence score is
\begin{equation}
\label{eq:permanence}
\mathrm{Perm}(p,C_i)=\frac{I(p,C_i)}{\max(E_{\max}(p,C_i),1)}-\left(1-C_{\mathrm{in}}(p,C_i)\right).
\end{equation}
High permanence indicates that $p$ has many neighbours inside $C_i$, few concentrated neighbours in a competing module, and internally cohesive neighbours. Low or negative permanence highlights boundary proteins whose assignment deserves inspection.

\begin{algobox}{Permanence calculation for one protein--module pair}
\label{alg:permanence}
\item Input: protein $p$, module $C_i$, graph $G$, and module set $\mathcal{C}$.
\item Compute the internal degree $I=|N(p)\cap C_i|$.
\item Compute $E_{\max}=\max_{C_j\neq C_i}|N(p)\cap C_j|$.
\item Set $C_{\mathrm{in}}=0$ when $I<2$; otherwise compute the clustering coefficient among $N(p)\cap C_i$.
\item Return $\mathrm{Perm}(p,C_i)$ from Eq.~\eqref{eq:permanence}.
\end{algobox}

\subsection{Functional dependency and membership score}
Functional dependency measures whether the annotations of protein $p$ align with the GO signature of module $C_i$:
\begin{equation}
\label{eq:fd}
\mathrm{fd}(p,C_i)=
\begin{cases}
\frac{1}{|G(p)|}\sum_{t\in G(p)} \mathrm{TFIDF}(t,C_i), & |G(p)|>0,\\
0, & |G(p)|=0.
\end{cases}
\end{equation}
The overlap seed combines structural and functional evidence as
\begin{equation}
\label{eq:membership}
M(p,C_i)=\alpha\,\mathrm{Perm}(p,C_i)+(1-\alpha)\,\mathrm{fd}(p,C_i),
\end{equation}
where $\alpha\in[0,1]$ balances topology and function. In the current ECHO-PPI implementation this membership formulation supports MCL+overlap seeding and audit records. It should be read as a transparent scoring rule, not as a guarantee that higher score always means experimentally confirmed membership.

\begin{algobox}{Functional dependency and membership score}
\label{alg:membership}
\item Input: protein $p$, module $C_i$, TF--IDF scores, permanence scores, and weight $\alpha$.
\item If $G(p)$ is empty, set $\mathrm{fd}(p,C_i)=0$.
\item Otherwise compute $\mathrm{fd}(p,C_i)$ from Eq.~\eqref{eq:fd}.
\item Compute $M(p,C_i)$ from Eq.~\eqref{eq:membership}.
\item Return $\mathrm{fd}(p,C_i)$ and $M(p,C_i)$.
\end{algobox}

\subsection{Overlap addition and transfer rules}
For a protein currently assigned to module set $\mathcal{C}_p$, ECHO-PPI evaluates the overlap-gain criterion
\begin{equation}
\label{eq:overlap-rule}
M(p,C_j\cup\{p\})-\max_{C_k\in\mathcal{C}_p}M(p,C_k)>\tau_{\mathrm{overlap}}.
\end{equation}
The additional assignment to $C_j$ is accepted only when this inequality holds. This rule permits biologically plausible multi-membership only when the candidate module improves the assignment score by a defined margin. A transfer rule handles probable misplacements: if extra-cluster links exceed intra-cluster links and the best external module $C_{\max}$ has stronger internal connectivity for $p$ than the current module, $p$ may be moved to $C_{\max}$. In ECHO-PPI this layer is used conservatively; its role is to expose and document overlap, not to inflate modules aggressively.
The gain criterion uses the existing TF--IDF signature of $C_j$ as an approximation; the contribution of $p$ to the module signature is not recomputed during evaluation.

\begin{algobox}{Overlap reassignment with transfer check}
\label{alg:overlap}
\item Input: initial modules $\mathcal{C}$, graph $G$, GO map, scores $M$, and threshold $\tau_{\mathrm{overlap}}$.
\item For each protein $p$, collect its current module set $\mathcal{C}_p=\{C_i:p\in C_i\}$.
\item For each candidate module $C_j\notin\mathcal{C}_p$, test the gain condition in Eq.~\eqref{eq:overlap-rule}.
\item If the gain exceeds $\tau_{\mathrm{overlap}}$, add $p$ to $C_j$ as an overlapping assignment.
\item For each current module $C_i\in\mathcal{C}_p$, compare extra-cluster and intra-cluster links.
\item If the best external module $C_{\max}$ has stronger support than $C_i$, transfer $p$ to $C_{\max}$.
\item Return the updated module set.
\end{algobox}

\subsection{Evidence-potential nucleus score}
Within ECHO-PPI, the black-hole metaphor denotes a deterministic evidence-potential mechanism for high-support local nuclei. \emph{Mass} is the nucleus score; \emph{attraction} is compatibility between a protein and a candidate community from topology, semantic similarity, and GO overlap; the confidence \emph{boundary} labels memberships without aggressive removal when all seed members are preserved.

\begin{equation}
\label{eq:bh}
\begin{aligned}
\mathrm{BH}(p)={}&
0.25\,\widehat{\mathrm{deg}}(p)
+0.15\,\widehat{\mathrm{cc}}(p)
+0.20\,\mathrm{kcore}(p)\\
&+0.20\,\mathrm{SNC}(p)
+0.20\,\widehat{\mathrm{rich}}(p)
-\lambda_{\mathrm{unc}}\,\widehat{\mathrm{unc}}(p),
\end{aligned}
\end{equation}
where $\lambda_{\mathrm{unc}}=0.05$. The positive evidence weights sum to 1.00, while $\widehat{\mathrm{unc}}(p)$ is treated as a separately scaled annotation-sparsity penalty. Here $\widehat{\mathrm{deg}}(p)$ is normalised weighted degree, $\widehat{\mathrm{cc}}(p)$ normalised clustering coefficient, $\mathrm{kcore}(p)$ normalised $k$-core score, $\mathrm{SNC}(p)$ semantic neighbourhood coherence (mean cosine similarity between $e_p$ and neighbour embeddings), $\widehat{\mathrm{rich}}(p)$ normalised non-generic GO richness, and $\widehat{\mathrm{unc}}(p)$ normalised annotation uncertainty. Nuclei are selected by greedy non-maximum suppression on $\mathrm{BH}(p)$ with minimum hop separation; they score candidates but do not alone define the final partition (see core-only ablation).
The evidence-potential nucleus features and their weights are summarised in Table~\ref{tab:bh-features}.

\begin{table}[t]
\centering
\caption{Evidence-potential nucleus features and weights.}
\label{tab:bh-features}
\scriptsize
\begin{tabular}{p{1.55cm}p{1.05cm}p{4.25cm}}
\toprule
Feature & Weight & Role \\
\midrule
$\widehat{\mathrm{deg}}(p)$ & 0.25 & Connectivity in weighted PPI graph \\
$\widehat{\mathrm{cc}}(p)$ & 0.15 & Local cohesion \\
$\mathrm{kcore}(p)$ & 0.20 & Core--periphery position \\
$\mathrm{SNC}(p)$ & 0.20 & Semantic coherence with neighbours \\
$\widehat{\mathrm{rich}}(p)$ & 0.20 & Non-generic GO evidence \\
$\widehat{\mathrm{unc}}(p)$ & penalty 0.05 & Annotation sparsity penalty \\
\bottomrule
\end{tabular}
\end{table}

\subsection{Candidate generation and scoring}
Candidates arise from MCL modules, nucleus-centred ego neighbourhoods (one--two hops), greedy topology--semantic expansion, semantic $k$-nearest neighbours with weak graph support, and hybrid unions when Jaccard overlap $\geq 0.5$. Typical pool size on the cleaned Gavin graph is $\sim 850$ modules. Each candidate receives a composite score combining topology cohesion, maximum $\mathrm{BH}$ among members, semantic coherence, GO coherence, size prior, uncertainty, and fragmentation penalties.

\begin{algobox}{Candidate community generation}
\label{alg:candidates}
\item Initialise $\mathcal{C}$ with MCL$(G)$ modules.
\item For each high-scoring nucleus $q$, add hop-limited ego modules around $q$.
\item Add greedy topology--semantic expansions around each nucleus.
\item Add semantic $k$-NN sets that pass the graph-support filter.
\item Add hybrid unions of candidates with sufficient Jaccard overlap.
\item Return the deduplicated candidate set $\mathcal{C}$.
\end{algobox}

\subsection{Overlap seeding and recall-safe supplementation}
Final modules are seeded from MCL with function-aware overlap reassignment (permanence and GO TF--IDF; inflation $2.0$). Recall-safe supplementation adds boundary proteins only if relative size increase $\leq 15\%$, at most two extra members per module, and evidence gain $\geq 0.38$ with OR-gated topology ($\geq 0.12$), semantic ($\geq 0.28$), or GO ($\geq 0.25$) support.

\begin{algobox}{Recall-safe supplementation}
\label{alg:supplementation}
\item For each seed module $M_0$, initialise $M=M_0$ and enumerate the boundary $\partial M$.
\item While the size cap is not reached and $\partial M$ is non-empty, rank boundary proteins by gated evidence gain.
\item Select the best boundary protein $p^*\in\partial M$.
\item If the gain is below threshold, stop expanding the current module.
\item Otherwise set $M=M\cup\{p^*\}$ and update $\partial M$.
\item Return the supplemented modules.
\end{algobox}

\subsection{Hierarchical confidence labels}
For each $(p,C)$ we compute topology support $\tau_p$ and semantic support $\sigma_p$. Labels are assigned as: \textbf{core} if $\tau_p\geq 0.35$ and $\sigma_p\geq 0.25$; \textbf{inner} if either $\geq 0.25$; \textbf{outer} if either $\geq 0.12$; otherwise \textbf{uncertain}. With preserve-all seeding, proteins are retained for audit even when uncertain.

The observed label distribution depends on the normalisation range and embedding cache used to score assignments. In the cleaned Gavin run, removal of spreadsheet-corrupted identifiers and rebuilding of the TF--IDF/SVD semantic cache shifted semantic support upward while the fixed label thresholds were unchanged. The resulting labels are therefore interpreted primarily as a Core-versus-non-Core triage signal; Inner assignments provide weaker evidence, while Outer and Uncertain assignments expose boundary cases for manual inspection. The hierarchical confidence-label rules are summarised in Table~\ref{tab:labels}.

\begin{table}[t]
\centering
\caption{Hierarchical confidence label rules.}
\label{tab:labels}
\small
\begin{tabular}{ll}
\toprule
Label & Rule \\
\midrule
Core & $\tau_p \geq 0.35$ and $\sigma_p \geq 0.25$ \\
Inner & $\tau_p \geq 0.25$ or $\sigma_p \geq 0.25$ \\
Outer & $\tau_p \geq 0.12$ or $\sigma_p \geq 0.12$ \\
Uncertain & otherwise \\
\bottomrule
\end{tabular}
\end{table}

\subsection{Evidence bundles and evaluation}
Each assignment exports protein and module identifiers, confidence label, topology, semantic, and GO scores, optional stability frequency, top non-generic GO terms, and a template evidence summary sentence. Table~\ref{tab:bundle-fields} lists the required evidence-bundle fields. Completeness of these fields is a documentation metric: it measures whether the assignment can be audited, not whether the assignment is biologically correct.

\begin{table}[t]
\centering
\caption{Assignment-level evidence-bundle schema. Required fields make each protein--module assignment inspectable; optional fields support stability and downstream review.}
\label{tab:bundle-fields}
\footnotesize
\begin{tabular}{p{3.0cm}p{8.8cm}}
\toprule
Field & Purpose \\
\midrule
\texttt{protein\_id}, \texttt{community\_id} & Assignment keys linking a protein to a predicted module \\
\texttt{membership\_type} & Hierarchical confidence label: core, inner, outer, or uncertain \\
\texttt{topology\_score} & Graph-neighbour support for this assignment \\
\texttt{semantic\_score} & Embedding or text-profile support for this assignment \\
\texttt{go\_score} & GO-term evidence channel, when non-generic annotations are available \\
\texttt{membership\_score} & Combined assignment strength used for ranking or triage \\
\texttt{top\_go\_terms} & Module-specific GO terms, typically ranked by TF--IDF \\
\texttt{evidence\_summary} & Short natural-language explanation for curator review \\
\texttt{stability\_freq.} & Optional frequency across seeds or parameter settings \\
\bottomrule
\end{tabular}
\end{table}

Evaluation reports a deterministic full-gold benchmark and five held-out gold-complex splits (seeds 42--46). We report F1, precision, recall, matched-protein coverage, mean and median module size where available, required-field evidence-bundle completeness, overlap rate, and runtime. Ablation configurations include score-select-only (candidate scoring without overlap seed), naive boundary expansion (negative control), and core-only nucleus seeding.

Baseline methods include MCL, MCL with function-aware overlap reassignment, exact ClusterONE 1.0 using the official Paccanaro laboratory JAR on the same cleaned weighted graph, and an SLPA sensitivity grid with thresholds 0.05, 0.10, 0.15, and 0.20 for 100 iterations. ClusterONE is reported as a main external overlapping baseline; SLPA is reported as sensitivity because its threshold choice strongly affects the number and size of overlapping modules. The same matching and auditability protocol is applied to Gavin and Krogan.

\begin{algobox}{Greedy Jaccard matching for F1}
\label{alg:jaccard}
\item Sort all predicted--gold pairs by $\mathrm{J}(C,G)$ in descending order.
\item Visit the sorted pairs sequentially.
\item If both modules are unmatched and $\mathrm{J}\geq 0.5$, count one true positive.
\item Mark the accepted predicted module and gold complex as matched.
\item Compute precision, recall, and F1 from the final match counts.
\end{algobox}

\section{Results}

\subsection{Candidate generation expands the search space but does not guarantee coverage}
\begin{figure}[htbp]
\centering
\IfFileExists{Figure_2.pdf}{%
\includegraphics[width=\linewidth]{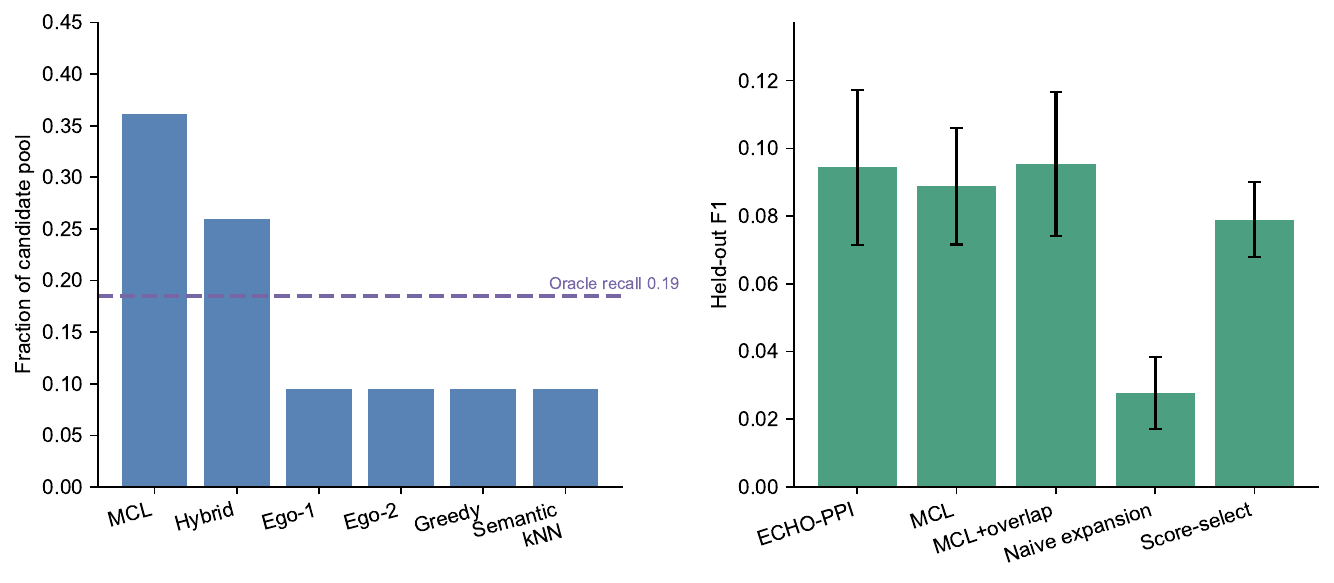}%
}{%
\fbox{\parbox{0.85\linewidth}{\centering Figure 2 file not found. Original author-approved figure file missing.}}%
}
\caption{Candidate-source composition and held-out benchmark summary (candidate oracle recall upper bound $=0.185$).}
\label{fig:candidate}
\end{figure}

The candidate pool contains 850 modules on the cleaned Gavin graph. Oracle analysis---selecting, for each gold complex, the candidate with highest Jaccard---shows that 18.5\% of gold complexes have a candidate with best Jaccard $\geq 0.5$, 54.6\% have partial candidate coverage below the matching threshold, and 26.9\% have no overlapping candidate at all (Table~\ref{tab:oracle}; Figure~\ref{fig:candidate}). The mean and median best-candidate Jaccard values are 0.246 and 0.156, respectively. This ceiling reflects limitations of candidate coverage, graph sparsity, complex size, and identifier/gold-standard compatibility, not only the final selection step.

\begin{table}[t]
\centering
\caption{Candidate-pool oracle analysis on Gavin.}
\label{tab:oracle}
\scriptsize
\begin{tabular}{p{2.25cm}rp{3.35cm}}
\toprule
Category & Fraction & Interpretation \\
\midrule
Best Jaccard $=0$ & 0.269 & No candidate overlap \\
$0<$ Best Jaccard $<0.5$ & 0.546 & Partial, below match threshold \\
Best Jaccard $\geq0.5$ & 0.185 & Matchable by oracle selection \\
\bottomrule
\end{tabular}
\end{table}

\subsection{Exact ClusterONE is the strongest predictive baseline on Gavin}
\begin{figure}[htbp]
\centering
\IfFileExists{Figure_3.pdf}{%
\includegraphics[width=\linewidth]{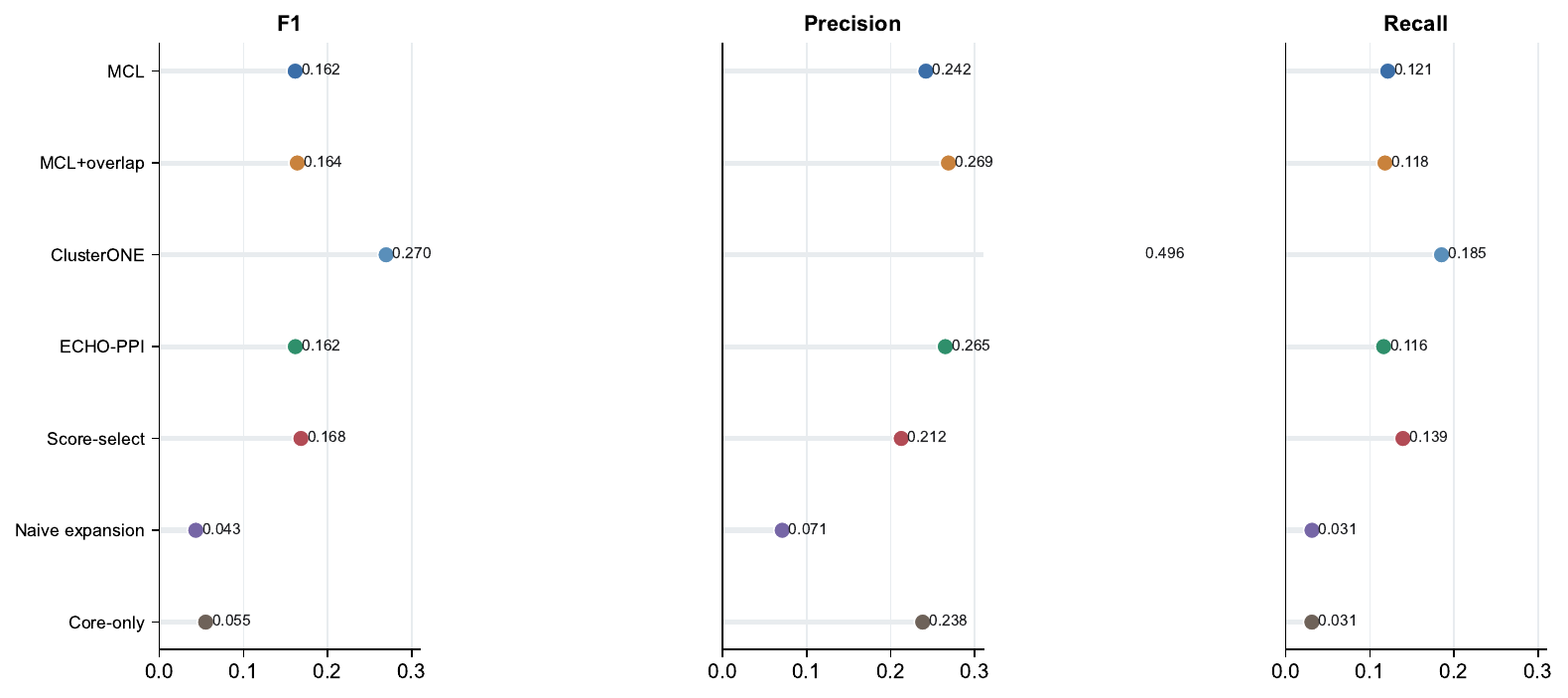}%
}{%
\fbox{\parbox{0.85\linewidth}{\centering Figure 3 file not found. Original author-approved figure file missing.}}%
}
\caption{Gavin benchmark comparison across F1, precision, and recall. ClusterONE is an exact external baseline; ECHO-PPI trails it predictively while adding evidence bundles.}
\label{fig:bench}
\end{figure}

On the full gold standard (Table~\ref{tab:bench-full}; Figure~\ref{fig:bench}), exact ClusterONE achieves the strongest predictive result (F1 $=0.270$, precision $=0.496$, recall $=0.185$). MCL, MCL+overlap, and ECHO-PPI are close to one another (F1 $=0.162$, $0.164$, and $0.162$, respectively). ECHO-PPI therefore does \textbf{not} claim predictive superiority on Gavin; its measurable contribution is the assignment-level audit layer. Table~\ref{tab:auditability} summarises this distinction by comparing predictive performance, overlap support, evidence fields, and confidence labels across the main methods.

\begin{table}[t]
\centering
\caption{Gavin full gold-standard benchmark.}
\label{tab:bench-full}
\small
\resizebox{\textwidth}{!}{%
\begin{tabular}{lrrrrr}
\toprule
Method & F1 & Precision & Recall & Mean size & Bundle compl. \\
\midrule
MCL & 0.162 & 0.242 & 0.121 & 5.98 & 0.00 \\
MCL + overlap & 0.164 & 0.269 & 0.118 & 6.91 & 0.00 \\
ClusterONE & 0.270 & 0.496 & 0.185 & 5.11 & 0.00 \\
SLPA ($t=0.05$) & 0.131 & 0.264 & 0.087 & 12.45 & 0.00 \\
\textbf{ECHO-PPI} & \textbf{0.162} & \textbf{0.265} & \textbf{0.116} & \textbf{7.13} & \textbf{1.00} \\
Score-select ablation & 0.168 & 0.213 & 0.139 & 13.40 & 1.00 \\
Naive expansion control & 0.043 & 0.071 & 0.031 & 39.41 & 1.00 \\
Core-only ablation & 0.055 & 0.238 & 0.031 & 12.8 & 0.00 \\
\addlinespace
\multicolumn{6}{p{0.92\textwidth}}{\footnotesize SLPA performance varies substantially with threshold choice; the reported row corresponds to the best F1 configuration from the sensitivity grid.} \\
\bottomrule
\end{tabular}
}%
\end{table}

\begin{figure}[htbp]
\centering
\IfFileExists{Figure_4.pdf}{%
\includegraphics[width=0.72\linewidth]{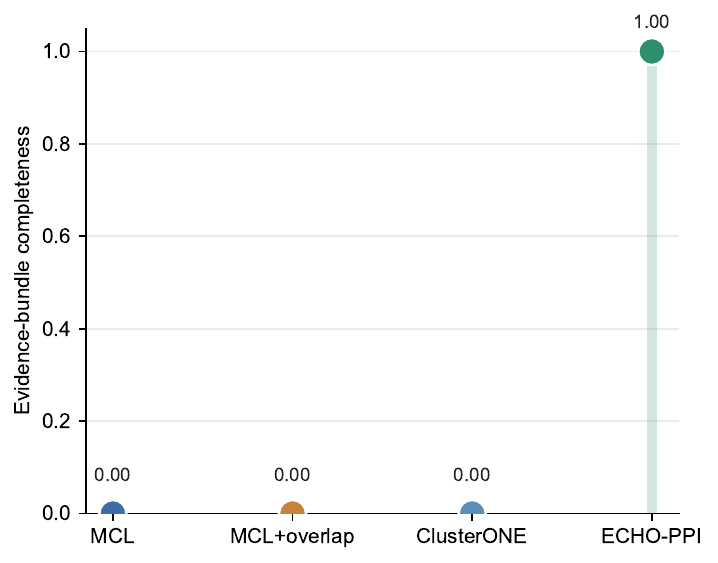}%
}{%
\fbox{\parbox{0.85\linewidth}{\centering Figure 4 file not found. Original author-approved figure file missing.}}%
}
\caption{Required-field evidence-bundle completeness. ECHO-PPI exports complete assignment-level evidence fields; hard baselines export no per-assignment bundles.}
\label{fig:audit}
\end{figure}

\begin{table}[t]
\centering
\caption{Auditability comparison. ClusterONE is stronger predictively on Gavin but does not export assignment-level evidence bundles or confidence labels.}
\label{tab:auditability}
\small
\resizebox{\textwidth}{!}{%
\begin{tabular}{lcccc}
\toprule
Method & F1 & Overlap output & Evidence fields & Confidence labels \\
\midrule
MCL & 0.162 & No & No & No \\
MCL + overlap & 0.164 & Limited & No & No \\
ClusterONE & 0.270 & Yes & No & No \\
\textbf{ECHO-PPI} & \textbf{0.162} & \textbf{Yes} & \textbf{Yes} & \textbf{Yes} \\
SLPA (sensitivity) & 0.131 & Yes & No & No \\
\bottomrule
\end{tabular}
}%
\end{table}

\subsection{Score-select and naive expansion reveal the precision--recall--size trade-off}
\begin{figure}[htbp]
\centering
\IfFileExists{Figure_5.pdf}{%
\includegraphics[width=0.85\linewidth]{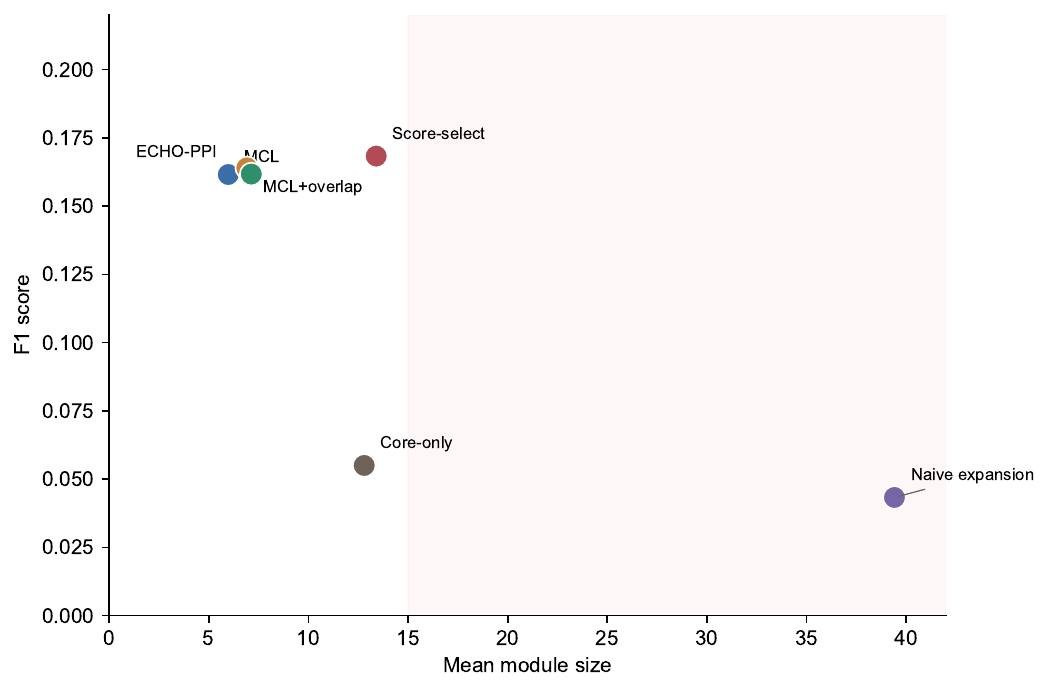}%
}{%
\fbox{\parbox{0.85\linewidth}{\centering Figure 5 file not found. Original author-approved figure file missing.}}%
}
\caption{Precision--recall--size trade-off. Naive expansion occupies a high-size, low-F1 region; ECHO-PPI remains near the overlap heuristic with modest size increase.}
\label{fig:tradeoff}
\end{figure}

\begin{figure}[htbp]
\centering
\IfFileExists{Figure_6.pdf}{%
\includegraphics[width=0.78\linewidth]{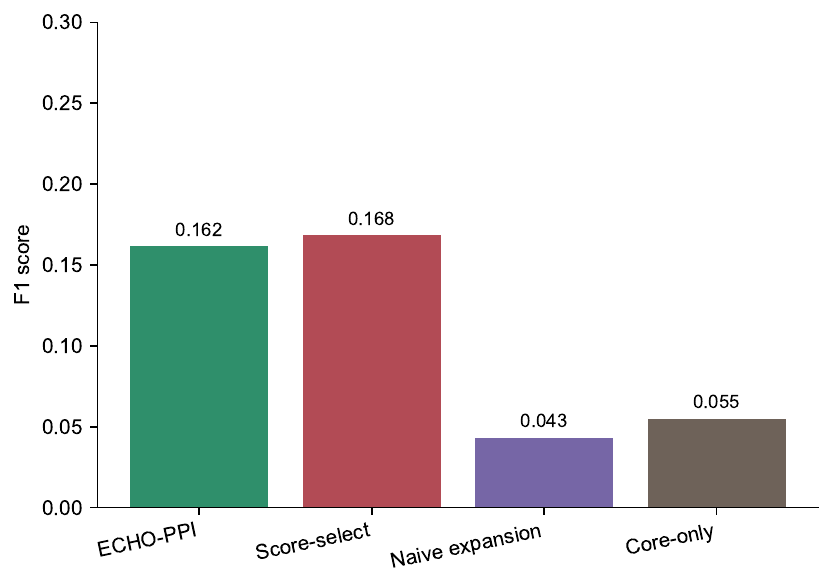}%
}{%
\fbox{\parbox{0.85\linewidth}{\centering Figure 6 file not found. Original author-approved figure file missing.}}%
}
\caption{Failure-mode ablation summary on full gold.}
\label{fig:failure}
\end{figure}

Score-select-only ablation raises recall to $0.139$ but remains below ClusterONE and produces larger modules (mean size $13.40$). Naive boundary expansion inflates mean size to $39.41$ and lowers F1 to $0.043$. Core-only nucleus seeding achieves high precision in the legacy diagnostic but recall remains $0.031$ (Figure~\ref{fig:tradeoff}, Figure~\ref{fig:failure}).

\subsection{Krogan benchmark tests transferability across PPI resources}
\begin{figure}[htbp]
\centering
\IfFileExists{Figure_7.pdf}{%
\includegraphics[width=\linewidth]{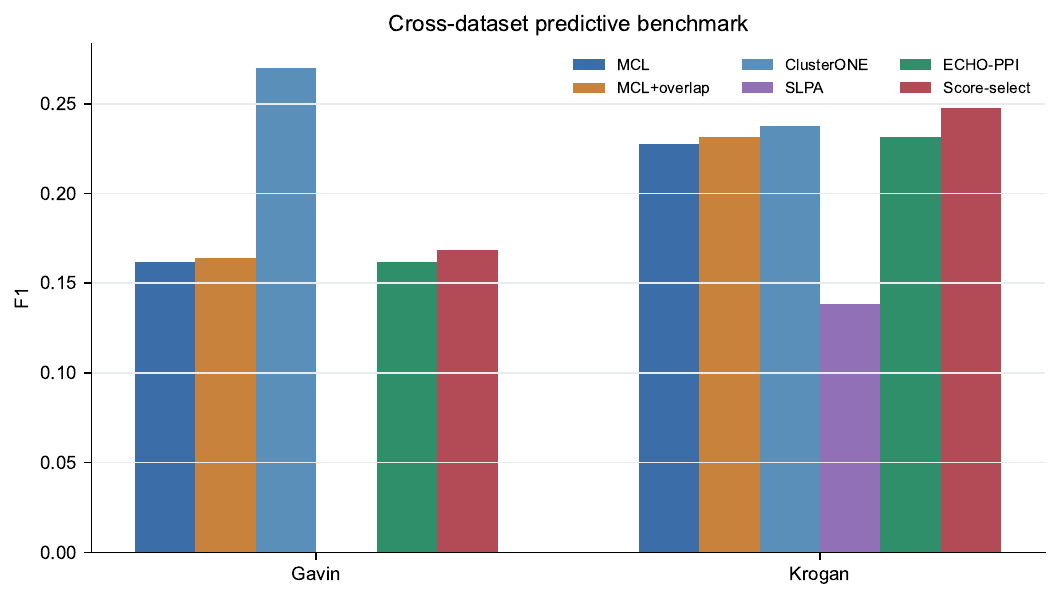}%
}{%
\fbox{\parbox{0.85\linewidth}{\centering Figure 7 file not found. Original author-approved figure file missing.}}%
}
\caption{Gavin and Krogan benchmark comparison. The second benchmark is used to test transferability of the workflow, not to tune for maximum F1.}
\label{fig:cross-benchmark}
\end{figure}

The Krogan Nature 2006 benchmark provides a second yeast PPI resource with different preprocessing and interaction characteristics. On this dataset, ECHO-PPI again matches the MCL+overlap seed predictively (F1 $=0.232$ for both; Table~\ref{tab:krogan-benchmark}; Figure~\ref{fig:cross-benchmark}). Exact ClusterONE is slightly stronger than ECHO-PPI (F1 $=0.237$), while score-select achieves the highest F1 in this fixed run (F1 $=0.248$), at the cost of substantially larger modules. These results do not support a predictive-superiority claim. Instead, they show that the evidence-bundle and confidence-label layer transfers to a second PPI resource while preserving the conservative interpretation established on Gavin.

\begin{table}[t]
\centering
\caption{Krogan 2006 full gold-standard benchmark using the same CYC2008 matching protocol.}
\label{tab:krogan-benchmark}
\small
\resizebox{\textwidth}{!}{%
\begin{tabular}{lrrrrr}
\toprule
Method & F1 & Precision & Recall & Mean size & Bundle compl. \\
\midrule
MCL & 0.227 & 0.230 & 0.225 & 4.43 & 0.00 \\
MCL + overlap & 0.232 & 0.247 & 0.218 & 5.06 & 0.00 \\
ClusterONE & 0.237 & 0.229 & 0.246 & 4.30 & 0.00 \\
SLPA best grid & 0.138 & 0.223 & 0.100 & 14.33 & 0.00 \\
\textbf{ECHO-PPI} & \textbf{0.232} & \textbf{0.247} & \textbf{0.218} & \textbf{5.23} & \textbf{1.00} \\
Score-select ablation & 0.248 & 0.313 & 0.205 & 9.47 & 1.00 \\
Naive expansion control & 0.030 & 0.032 & 0.028 & 40.79 & 1.00 \\
Core-only ablation & 0.157 & 0.675 & 0.089 & 9.99 & 0.00 \\
\bottomrule
\end{tabular}
}%
\end{table}

\subsection{Evidence bundles provide assignment-level interpretability}
ECHO-PPI achieves required-field evidence-bundle completeness of $1.00$, whereas MCL and MCL+overlap export no assignment-level bundles (Figure~\ref{fig:audit}). This metric measures documentation coverage, not biological correctness; evidence-channel scores may legitimately be zero for weakly supported assignments.

\begin{figure}[htbp]
\centering
\IfFileExists{Figure_8.pdf}{%
\includegraphics[width=\linewidth]{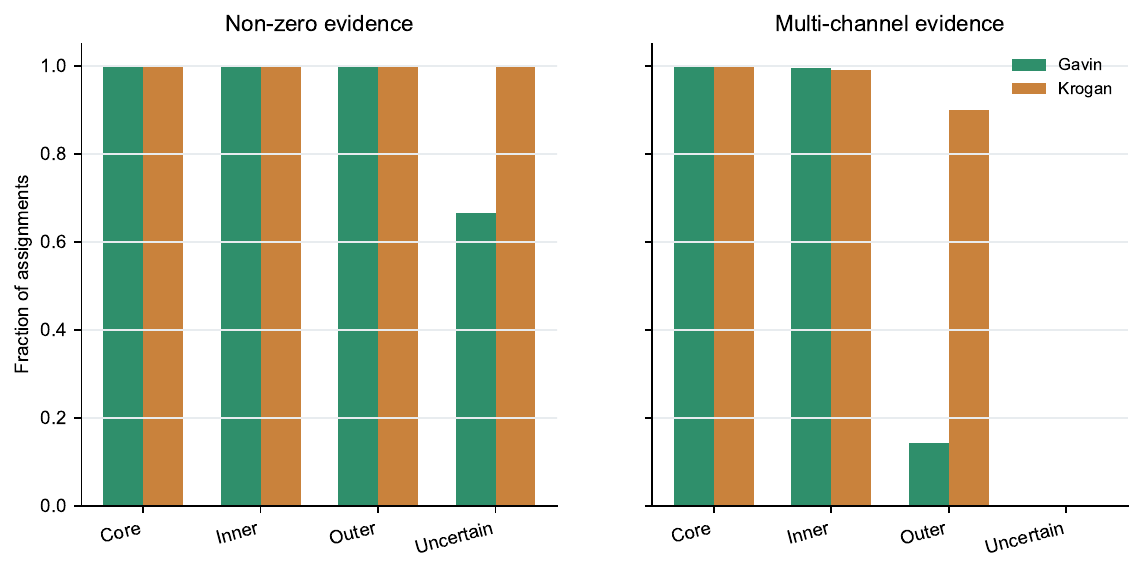}%
}{%
\fbox{\parbox{0.85\linewidth}{\centering Figure 8 file not found. Original author-approved figure file missing.}}%
}
\caption{Auditability transfer across Gavin and Krogan. Evidence-channel support remains high for Core assignments in both datasets, while Uncertain assignments are not promoted to multi-channel support.}
\label{fig:audit-transfer}
\end{figure}

Beyond field completeness, the generated support audit shows that all Gavin Core assignments contain non-zero evidence from topology, semantic, and GO channels, and all have support from at least two channels. Inner assignments are also usually multi-channel ($0.996$), whereas Uncertain assignments have lower non-zero evidence coverage ($0.667$) and no multi-channel support in the current Gavin run. On Krogan, Core assignments again show complete multi-channel support ($1.000$), Inner assignments remain mostly multi-channel ($0.991$), and Uncertain assignments have no multi-channel evidence (Figure~\ref{fig:audit-transfer}). These statistics are reported in \path{results/evidence_support_by_label.csv}, \path{results/krogan/auditability_summary.csv}, and \path{results/cross_dataset_auditability.csv}.

\subsection{Confidence labels identify high-support Core assignments}
\begin{figure}[htbp]
\centering
\IfFileExists{Figure_9.pdf}{%
\includegraphics[width=0.85\linewidth]{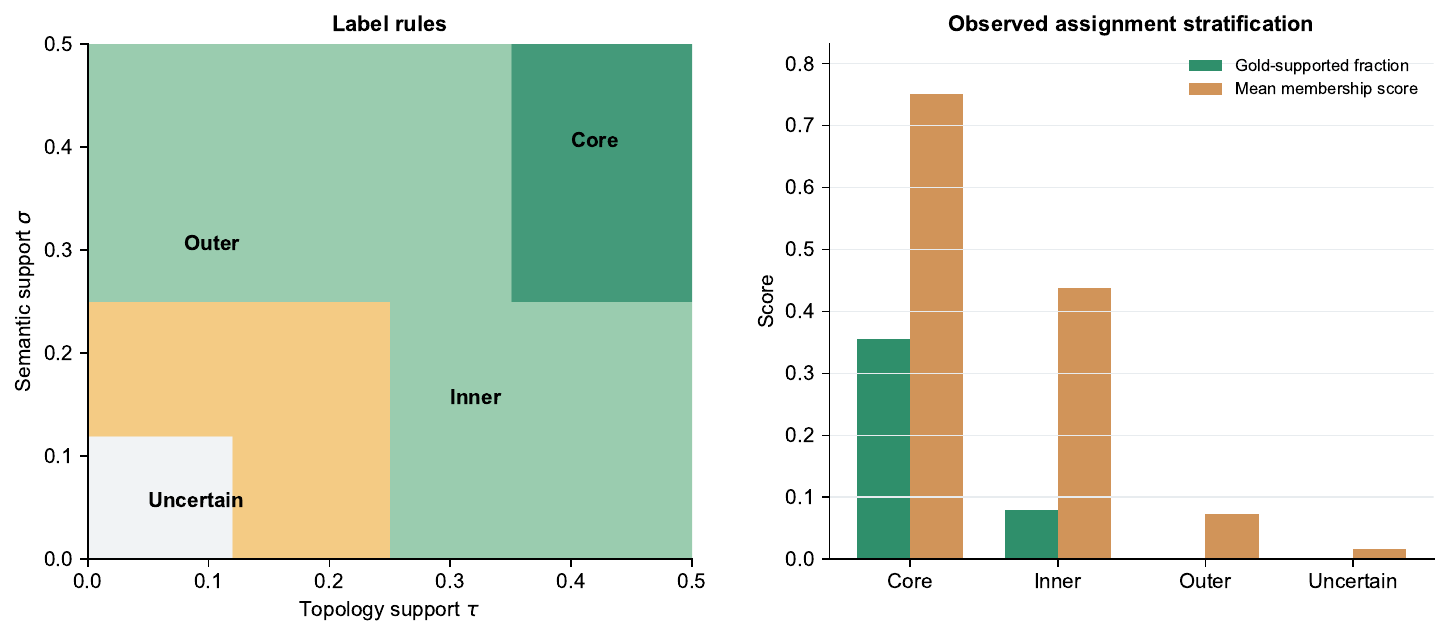}%
}{%
\fbox{\parbox{0.85\linewidth}{\centering Figure 9 file not found. Original author-approved figure file missing.}}%
}
\caption{Hierarchical confidence labels. Left: rule regions in topology--semantic support space. Right: observed assignment stratification by gold-supported fraction and membership score.}
\label{fig:labels}
\end{figure}

\begin{figure}[htbp]
\centering
\IfFileExists{Figure_10.pdf}{%
\includegraphics[width=0.82\linewidth]{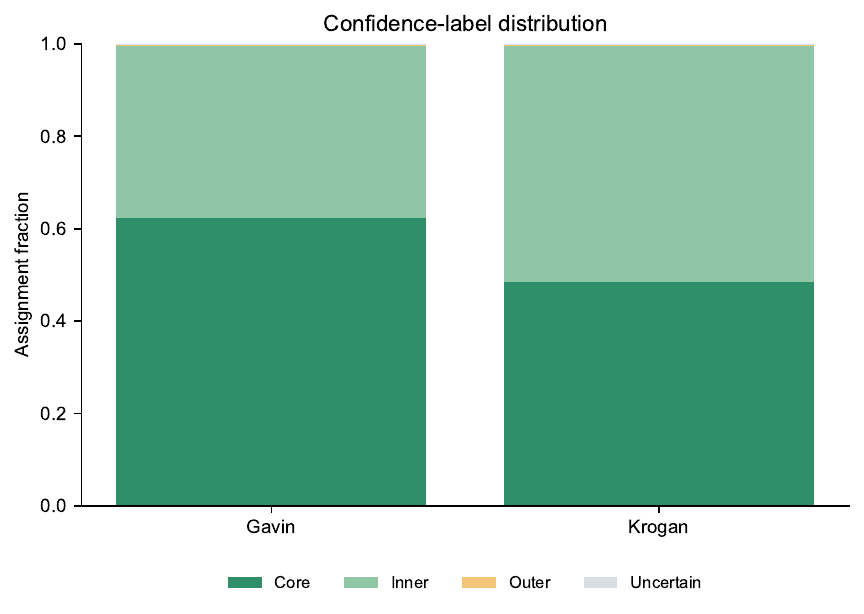}%
}{%
\fbox{\parbox{0.85\linewidth}{\centering Figure 10 file not found. Original author-approved figure file missing.}}%
}
\caption{Confidence-label distribution across Gavin and Krogan. Label proportions shift across resources, but the audit layer remains exportable and interpretable.}
\label{fig:label-transfer}
\end{figure}

ECHO-PPI confidence labels separate high-support assignments from weak boundary cases (Table~\ref{tab:label-validation}; Figure~\ref{fig:labels}; Figure~\ref{fig:label-transfer}). Core assignments have the highest mean membership score ($0.751$) and the highest gold-supported assignment fraction ($0.356$) on Gavin. Inner assignments show weaker gold support ($0.080$) and require manual inspection, whereas Outer and Uncertain assignments primarily expose boundary cases. The same pattern appears on Krogan: Core assignments have gold-supported fraction $0.347$, compared with $0.074$ for Inner and $0.000$ for Outer/Uncertain assignments. The current evidence supports a Core-versus-non-Core triage distinction rather than a fully calibrated four-level gradient.

\begin{table}[t]
\centering
\caption{Assignment-level validation of ECHO-PPI confidence labels. Gold-supported fraction is the fraction of protein--module assignments for which the predicted module matches a gold complex containing that protein at Jaccard $\geq 0.5$.}
\label{tab:label-validation}
\scriptsize
\resizebox{\textwidth}{!}{%
\begin{tabular}{lrrrrrr}
\toprule
Label & Assignments & Membership & Best gold Jaccard & Gold-supported frac. & Non-zero evidence frac. & Multi-ch. frac. \\
\midrule
Core & 1192 & 0.751 & 0.310 & 0.356 & 1.000 & 1.000 \\
Inner & 710 & 0.438 & 0.115 & 0.080 & 1.000 & 0.996 \\
Outer & 7 & 0.073 & 0.000 & 0.000 & 1.000 & 0.143 \\
Uncertain & 3 & 0.017 & 0.111 & 0.000 & 0.667 & 0.000 \\
\addlinespace
\multicolumn{7}{p{0.92\textwidth}}{\footnotesize Multi-channel: assignment has non-zero evidence score in at least two of the topology, semantic, and GO channels.} \\
\bottomrule
\end{tabular}
}%
\end{table}

\subsection{Held-out evaluation confirms that predictive superiority is not the ECHO-PPI claim}
On 20\% held-out reference complexes per seed, ClusterONE remains the strongest predictive method (F1 $=0.171\pm0.037$). ECHO-PPI remains close to MCL and MCL+overlap but does not exceed them materially (Table~\ref{tab:test}). Wilcoxon signed-rank tests over five held-out splits found no significant difference between ECHO-PPI and MCL+overlap ($p=0.317$); comparisons are exploratory because $N=5$. The ECHO-PPI versus ClusterONE comparison was also exploratory ($p=0.063$) and favoured ClusterONE numerically.

\begin{table}[t]
\centering
\caption{Held-out test benchmark (mean $\pm$ standard deviation over five gold-complex splits).}
\label{tab:test}
\small
\begin{tabular}{lrrr}
\toprule
Method & F1 & Precision & Recall \\
\midrule
ClusterONE & $0.171\pm0.037$ & $0.132\pm0.028$ & $0.246\pm0.053$ \\
MCL & $0.089\pm0.017$ & $0.062\pm0.012$ & $0.156\pm0.030$ \\
MCL + overlap & $0.095\pm0.021$ & $0.069\pm0.016$ & $0.152\pm0.034$ \\
\textbf{ECHO-PPI} & $\textbf{0.094}\pm\textbf{0.023}$ & $\textbf{0.069}\pm\textbf{0.017}$ & $\textbf{0.151}\pm\textbf{0.036}$ \\
Naive expansion control & $0.028\pm0.011$ & $0.020\pm0.008$ & $0.044\pm0.017$ \\
Score-select ablation & $0.079\pm0.011$ & $0.052\pm0.007$ & $0.169\pm0.024$ \\
\bottomrule
\end{tabular}
\end{table}

\subsection{Parameter sensitivity highlights the strength of tuned simple baselines}
An expanded local sensitivity analysis evaluated MCL inflation, SLPA thresholds, and ECHO-PPI supplementation thresholds on Gavin (Table~\ref{tab:sensitivity}). Default MCL inflation $2.0$ gives F1 $=0.162$, while inflation $3.0$ reaches F1 $=0.213$. SLPA is reproducible but weaker in this grid (best F1 $=0.131$ at threshold $0.05$). ECHO-PPI supplementation thresholds in the tested range ($0.30$--$0.46$ minimum evidence score) remain near F1 $=0.16$. This result reinforces the conservative interpretation: major predictive gains would require a stronger seed partition or candidate-selection mechanism, whereas ECHO-PPI's current value is the evidence layer around assignments.

\begin{table}[t]
\centering
\caption{Selected parameter sensitivity results on Gavin. These runs are used to interpret model behaviour, not to replace the main fixed-configuration benchmark.}
\label{tab:sensitivity}
\small
\begin{tabular}{llrrr}
\toprule
Parameter & Value & F1 & Precision & Recall \\
\midrule
MCL inflation & 1.4 & 0.123 & 0.281 & 0.079 \\
MCL inflation & 2.0 & 0.162 & 0.242 & 0.121 \\
MCL inflation & 3.0 & 0.213 & 0.267 & 0.177 \\
SLPA threshold & 0.05 & 0.131 & 0.264 & 0.087 \\
SLPA threshold & 0.20 & 0.128 & 0.320 & 0.080 \\
ECHO min evidence & 0.38 & 0.162 & 0.265 & 0.116 \\
ECHO min evidence & 0.46 & 0.164 & 0.269 & 0.118 \\
\bottomrule
\end{tabular}
\end{table}

\subsection{Secondary datasets require dedicated preprocessing}
STRING and BioGRID loaders are retained for reproducibility and future work, but their preliminary feasibility numbers are not used as main scientific results. The secondary networks differ strongly in edge semantics, density, identifier conventions, and compatibility with CYC2008 matching. Appendix~\ref{app:secondary-feasibility} records these implementation constraints without presenting broken preprocessing as evidence about biological performance.

\subsection{Runtime}
\begin{figure}[htbp]
\centering
\IfFileExists{Figure_11.pdf}{%
\includegraphics[width=0.78\linewidth]{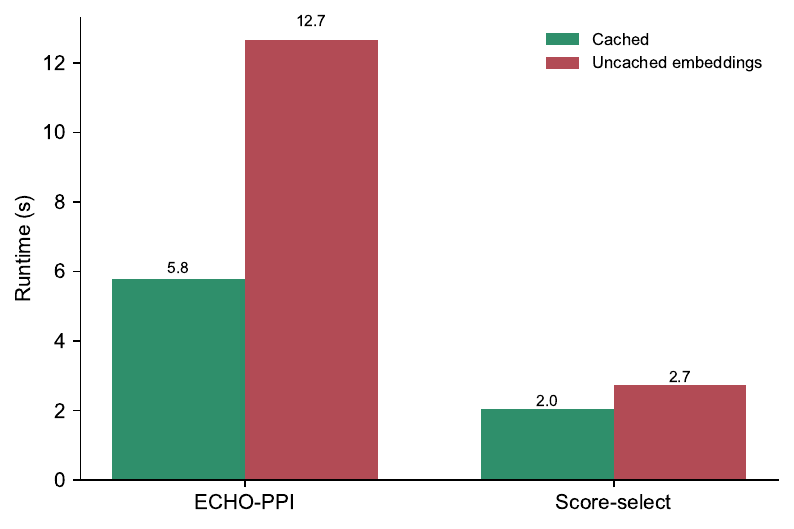}%
}{%
\fbox{\parbox{0.85\linewidth}{\centering Figure 11 file not found. Original author-approved figure file missing.}}%
}
\caption{Cached and uncached runtime comparison. Cached times assume pre-computed embeddings; uncached times include TF--IDF/SVD inference.}
\label{fig:runtime}
\end{figure}

With cached TF--IDF+SVD embeddings, ECHO-PPI requires approximately $5.78$\,s in the local Python environment; clearing the embedding cache raises this to $12.66$\,s. Score-select requires $2.05$\,s cached and $2.74$\,s with embeddings rebuilt (Figure~\ref{fig:runtime}). MCL and MCL+overlap are not embedding-dependent. Absolute runtimes depend on whether an external MCL binary or the Python fallback is used.

\subsection{Case studies illustrate curator-facing multi-membership evidence}
\begin{figure}[htbp]
\centering
\IfFileExists{Figure_12.pdf}{%
\includegraphics[width=0.82\linewidth]{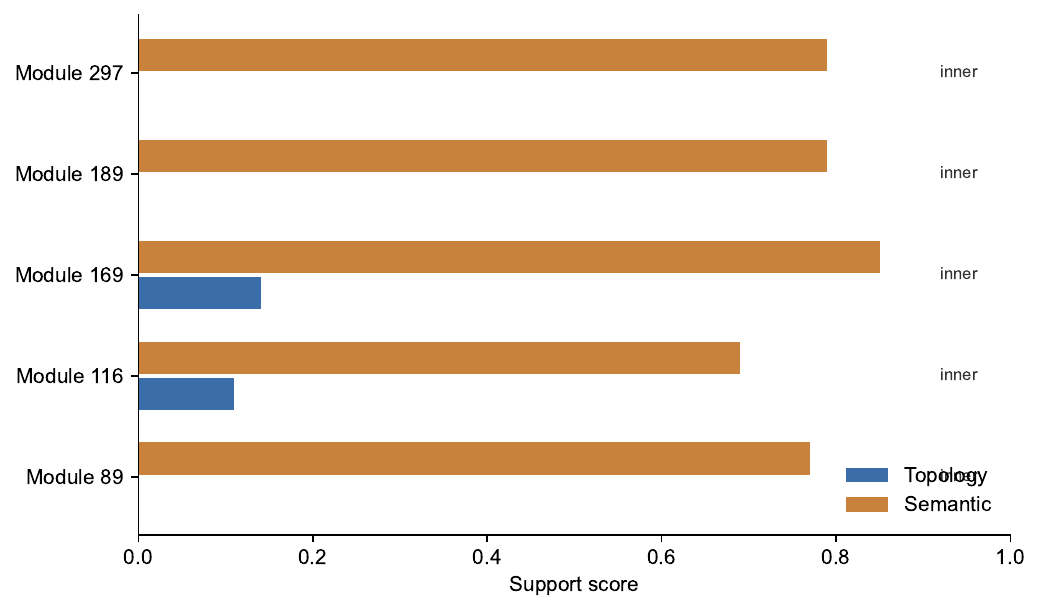}%
}{%
\fbox{\parbox{0.85\linewidth}{\centering Figure 12 file not found. Original author-approved figure file missing.}}%
}
\caption{Case study YKR018C: topology and semantic support across five modules (inner labels).}
\label{fig:case1}
\end{figure}

\begin{figure}[htbp]
\centering
\IfFileExists{Figure_13.pdf}{%
\includegraphics[width=0.82\linewidth]{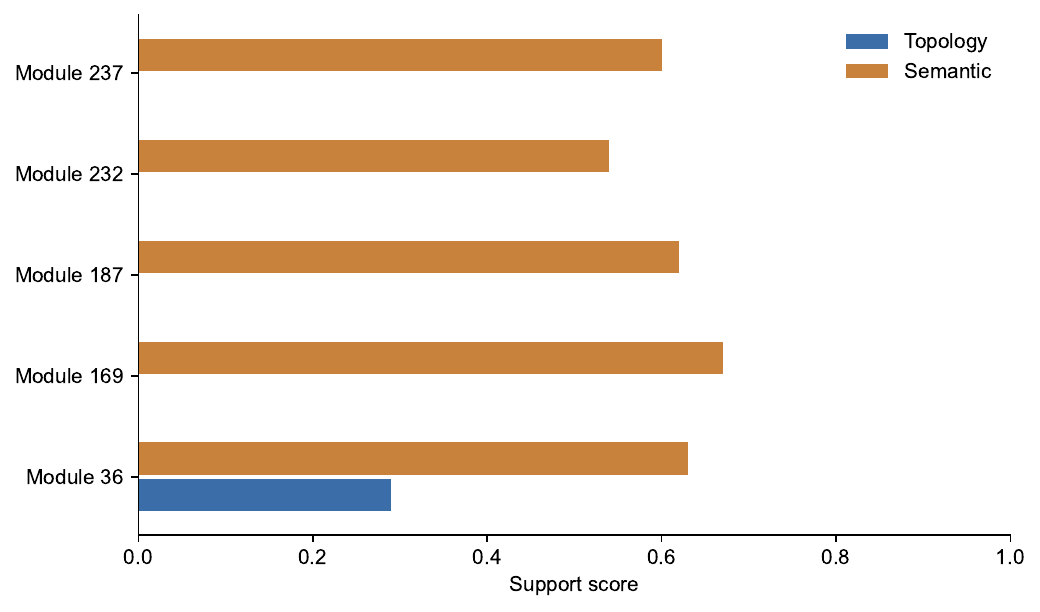}%
}{%
\fbox{\parbox{0.85\linewidth}{\centering Figure 13 file not found. Original author-approved figure file missing.}}%
}
\caption{Case study YIL161W: five modules with cytoplasmic GO support (inner labels).}
\label{fig:case2}
\end{figure}

Proteins \textbf{YKR018C} and \textbf{YIL161W} illustrate multi-module assignments with inner confidence labels and strong semantic support despite weak topology in some modules (Figure~\ref{fig:case1}; Figure~\ref{fig:case2}). These are algorithmic hypotheses for manual review, not experimentally validated complex memberships.

\begin{figure}[htbp]
\centering
\IfFileExists{Figure_14.pdf}{%
\includegraphics[width=\linewidth]{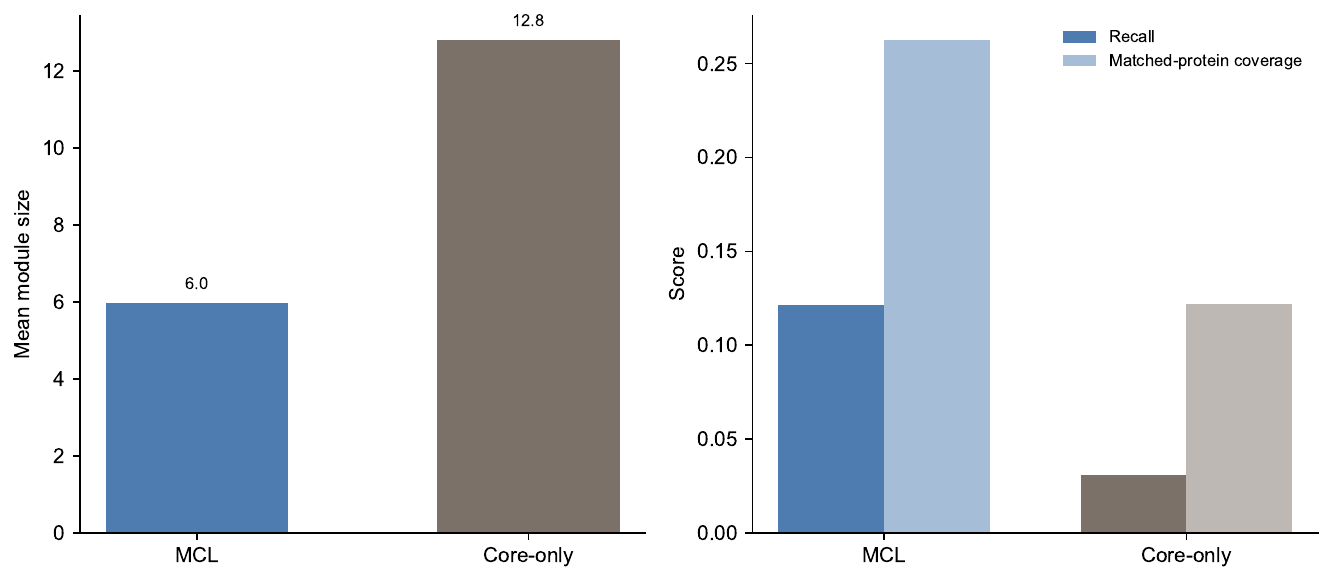}%
}{%
\fbox{\parbox{0.85\linewidth}{\centering Figure 14 file not found. Original author-approved figure file missing.}}%
}
\caption{Diagnostic comparison of module sizes and gold-protein coverage for MCL versus core-only ablation.}
\label{fig:diag}
\end{figure}

\begin{figure}[htbp]
\centering
\IfFileExists{Figure_15.pdf}{%
\includegraphics[width=0.72\linewidth]{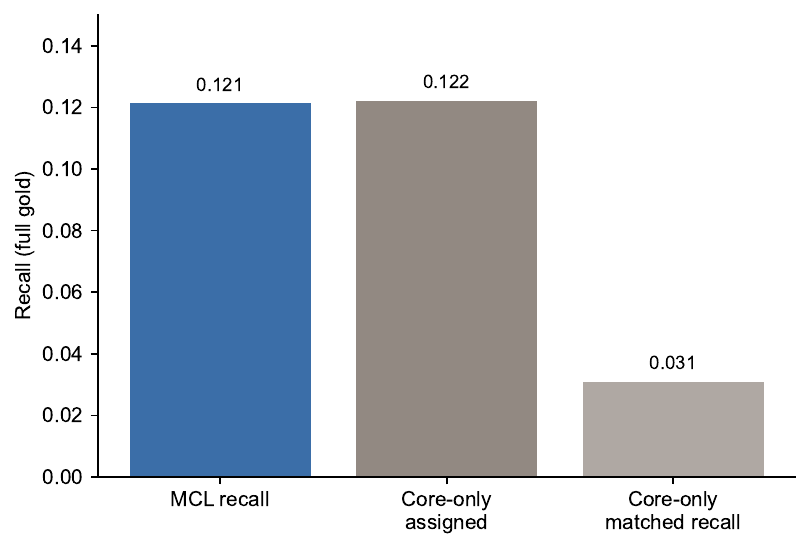}%
}{%
\fbox{\parbox{0.85\linewidth}{\centering Figure 15 file not found. Original author-approved figure file missing.}}%
}
\caption{Recall on full gold: MCL versus core-only ablation.}
\label{fig:recall-loss}
\end{figure}

Core-only ablation assigns far fewer gold proteins than MCL, explaining its recall collapse (Figure~\ref{fig:diag}; Figure~\ref{fig:recall-loss}).

\section{ECHO-PPI in practice: assignment-level walkthrough}

To make the workflow concrete, consider the case-study protein \textbf{YKR018C}. A hard MCL partition assigns this protein to a single module. ECHO-PPI instead treats the MCL result as a conservative seed, then asks whether additional protein--module assignments are supported by topology, semantic annotation, or GO evidence. The output is therefore not just ``YKR018C belongs to these modules'', but a set of auditable hypotheses with a confidence label and the evidence channel that made each assignment plausible.

In the first step, the weighted Gavin graph provides the local neighbourhood and the overlap seed. In the second step, ECHO-PPI builds the evidence profile for YKR018C, including graph-neighbour support, non-generic GO evidence, and an embedding-based semantic profile. In the third step, candidate modules near the seed and candidate modules semantically compatible with YKR018C are scored. For this protein, ECHO-PPI exports five inner-label assignments (Table~\ref{tab:scenario-ykr018c}; Figure~\ref{fig:case1}). The topology support is weak for several modules, but the semantic support is strong ($0.69$--$0.85$), with GO evidence including nuclear and cytoplasmic cellular-component terms (GO:0005634 and GO:0005737).

\begin{table}[t]
\centering
\caption{Worked ECHO-PPI assignment scenario for YKR018C. The assignments are hypotheses for review; the label records support strength rather than experimental proof.}
\label{tab:scenario-ykr018c}
\small
\begin{tabular}{lrrl}
\toprule
Assigned module & Topology support & Semantic support & Confidence label \\
\midrule
89  & 0.00 & 0.77 & Inner \\
116 & 0.11 & 0.69 & Inner \\
169 & 0.14 & 0.85 & Inner \\
189 & 0.00 & 0.79 & Inner \\
297 & 0.00 & 0.79 & Inner \\
\bottomrule
\end{tabular}
\end{table}

The key point is how the result should be read. ECHO-PPI does not silently convert all five assignments into equally confident complex memberships. Instead, each row remains inspectable: a curator can see that YKR018C is mostly supported by semantic evidence rather than dense local topology in these modules. If a downstream analysis needs conservative modules, it can keep only core or high-topology assignments; if the goal is hypothesis generation, it can prioritise inner assignments with strong semantic evidence for manual review. This scenario is the intended functional role of ECHO-PPI: preserving the competitive MCL+overlap prediction layer while adding the evidence ledger needed to decide which overlapping assignments are biologically persuasive, weak, or uncertain.

\subsection{Alternative overlapping-protein scenario}
YIL161W provides a second, complementary pattern. It is assigned to five modules with cytoplasmic GO support (GO:0005737) and inner confidence labels. One module has moderate topology support ($0.29$), while the other four are driven primarily by semantic support ($0.54$--$0.67$). For a curator, this distinction matters: the topology-supported assignment can be reviewed as a stronger local-network hypothesis, whereas the semantic-only assignments should be interpreted as weaker functional hypotheses that may reflect annotation similarity, shared cellular compartment, or missing PPI edges. The confidence label does not replace expert review; it makes the review queue more transparent.

A toy scenario can illustrate how permanence, GO TF--IDF, functional dependency, overlap gain, and transfer checks interact. ECHO-PPI keeps this decision logic but adds explicit evidence exports. A toy overlap would be accepted when $M(p,C_2\cup\{p\})-\max_{C_k\in\mathcal{C}_p}M(p,C_k)>\tau_{\mathrm{overlap}}$; a real ECHO-PPI record additionally reports whether that gain came from topology, semantic support, GO evidence, or a mixture. This separation prevents a high semantic score from being mistaken for direct physical evidence.

\section{Implementation and reproducibility}

The implementation is organised into reusable loading, evidence construction, community generation, evaluation, and plotting components. These details are intentionally kept in the repository documentation rather than in the main manuscript; the scientific requirement is that each reported table and figure can be regenerated from versioned scripts and machine-readable outputs.

To reproduce the fixed Gavin and Krogan analyses, install the project dependencies, place or symlink the Gavin PPI network, the BioGRID TAB3 source used for Krogan extraction, GO annotations, and CYC2008 gold standard under \texttt{data/}, and run:
\begin{verbatim}
export PYTHONPATH="$(pwd):$PYTHONPATH"
bash scripts/reproduce_echo_ppi_final.sh
\end{verbatim}
This command regenerates the Gavin primary benchmark, the Krogan transferability benchmark, benchmark tables, label-validation tables, auditability summaries, figures, and the manuscript PDF. The Krogan benchmark can also be regenerated directly with \texttt{python3} \path{scripts/evaluate_krogan_benchmark.py}; the repository README provides the module map, expected CSV schemas, evidence-bundle fields, and data-use notes needed for independent reproduction.

\section{Discussion}

ECHO-PPI adds auditability to overlapping PPI community detection but does not improve curated-complex F1 over strong predictive baselines on Gavin. Exact ClusterONE is substantially stronger on the current F1 benchmark, and tuned MCL can also exceed the default ECHO-PPI configuration. This result is scientifically important: predictive complex recovery and assignment-level auditability are related but distinct objectives. ECHO-PPI therefore preserves a conservative overlap-aware prediction layer and invests complexity in transparent assignment records.

The Krogan second benchmark strengthens the transferability claim without changing the predictive interpretation. The evidence-bundle and confidence-label workflow runs end-to-end on a different yeast interaction resource, produces complete assignment-level audit fields, and preserves the Core-versus-non-Core enrichment pattern. Predictively, however, ECHO-PPI remains close to MCL+overlap rather than dominating the strongest alternatives; score-select and ClusterONE can be numerically stronger depending on the dataset and objective.

The framework remains useful because module discovery is not only prediction. Curators and reviewers often need to know \emph{why} a protein appears in a module, whether the assignment is peripheral, and which GO or semantic signals support overlap. Evidence bundles and hierarchical confidence labels make these questions answerable at scale. Recall-safe supplementation further shows that bounded boundary expansion is preferable to naive growth: the latter increases module size without recall benefit and sharply reduces F1.

The assignment-level label analysis provides a more specific empirical basis for this auditability claim. Core assignments show the strongest enrichment for gold-supported memberships. Inner assignments provide weaker evidence and require manual inspection, whereas Outer and Uncertain assignments primarily expose boundary cases. The labels are therefore useful as triage metadata, especially for separating Core from non-Core assignments, but they should not be read as a fully calibrated four-tier probability scale.

Evidence-potential nuclei are valuable for ranking candidates and interpreting local structure, but insufficient as the sole seeding mechanism. Core-only ablation demonstrates that nucleus-centric partitions sacrifice recall even when precision remains high. Semantic embeddings provide functional priors yet depend on annotation completeness; graph filters remain essential.

Practically, ECHO-PPI is appropriate when assignment-level audit trails matter---for example, manual review pipelines, methods papers emphasising interpretability, or exploratory overlap analysis. It is especially useful when users need to rank protein--module assignments by confidence, separate strong local topology from semantic hypotheses, and export evidence records for downstream curation. When only maximum F1 against a fixed gold standard is required, tuned MCL or MCL+overlap may be the simpler and stronger choice.

MCL and MCL+overlap remain strong deterministic baselines, while ClusterONE and SLPA represent established overlapping community-detection families. The present results do not support a claim that ECHO-PPI dominates predictive F1. Instead, they support a narrower and more defensible claim: predictive parity with simple overlap-aware baselines plus assignment-level auditability through evidence bundles, confidence labels, and exportable support channels. Population-based optimisation of evidence weights and overlap thresholds is left for future work.

Future work should compare ECHO-PPI against additional exact overlapping baselines (OSLOM, BIGCLAM and link communities), evaluate STRING-scale and human PPI resources such as BioPlex with CORUM-compatible mapping, incorporate structural or cross-species evidence, and conduct curator-in-the-loop validation of evidence bundles.

\section{Limitations and outlook}

The Gavin benchmark shows no F1 improvement over MCL+overlap, and exact ClusterONE and tuned MCL can exceed the present ECHO-PPI F1; the contribution is auditability, not predictive leadership on this gold standard. On Krogan, ECHO-PPI again matches the overlap-seeded configuration and remains below score-select in F1. Complex Portal or CYC2008 mapping, GO identifier harmonisation, fallback MCL implementation, inflation tuning, and database version affect numerical values, and small differences between preprocessing snapshots should not be over-interpreted as algorithmic gains.

Additional external overlapping baselines and human PPI datasets remain to be tested. The local feasibility notes for STRING and BioGRID show that large networks require dataset-specific preprocessing and careful gold-standard harmonisation. Semantic components depend on annotation quality; missing GO annotations reduce evidence-channel support but do not imply absence of function. The evidence-potential score is a metaphorical ranking function, not a physical model. Case studies illustrate export format and multi-membership patterns but are not experimental validation.

Outlook includes genome-scale feasibility studies, uncertainty calibration of confidence labels, integration of structural interfaces, and systematic curator studies accepting or rejecting bundle-level hypotheses.

\section{Conclusion}

ECHO-PPI contributes a reproducible audit layer for overlapping protein--module assignments. Across Gavin and Krogan, ECHO-PPI remains close to MCL+overlap and adds hierarchical confidence labels and assignment-level evidence bundles. By combining topology, semantic and GO evidence with exportable records, ECHO-PPI makes module outputs inspectable at the level required for curator-facing workflows. Future work will further strengthen the predictive layer while preserving this assignment-level transparency.

\appendix
\section{Parameter reference}
\label{app:params}
Primary defaults: MCL inflation $2.0$; overlap blending $\alpha=0.5$, reassignment threshold $0.1$; supplementation relative size cap $15\%$, at most two added members, minimum evidence gain $0.38$; confidence thresholds as in Table~\ref{tab:labels}; Jaccard match threshold $0.5$; nucleus weights in Table~\ref{tab:bh-features}.

\section{Evidence bundle schema}
\label{app:bundle}
Each record includes protein identifier, community identifier, confidence label, topology, semantic, and GO scores, optional stability frequency, top non-generic GO terms, and a natural-language evidence summary.

\section{Additional ablation details}
\label{app:ablation}
Score-select-only chooses modules by candidate score without overlap seeding (F1 $0.168$, recall $0.139$). Naive expansion uses unbounded boundary growth (F1 $0.043$). Core-only ablation seeds communities from nuclei alone (F1 $0.055$, recall $0.031$).

\section{Checklist for new PPI datasets}
\label{app:checklist}
Researchers should harmonise identifiers; document gold-standard mapping; run MCL and overlap baselines; inspect candidate-pool oracle bounds; tune supplementation on validation complexes only; export bundles for manual QC; and report full-gold and held-out metrics separately.

\section{Secondary dataset feasibility notes}
\label{app:secondary-feasibility}
STRING and BioGRID loaders are available in the repository, but the present paper does not treat their preliminary local runs as main benchmarks. STRING is much denser than the Gavin socioaffinity graph and uses integrated association scores rather than direct complex-enrichment edge semantics. BioGRID requires additional evidence-type filtering, duplicate/self-loop removal, confidence handling, and careful alias harmonisation before direct CYC2008 matching is meaningful. These datasets should be promoted to secondary benchmarks only after dataset-specific preprocessing and gold-standard mapping are validated.

\backmatter

\bmhead{Data and code availability}
Source code, analysis scripts, processed result tables, figure-generation scripts, and documentation for reproducing the ECHO-PPI analyses are available at \url{https://github.com/MehrdadJalali-AI/ECHO-PPI}. Public input datasets were obtained from the Gavin yeast interaction study \cite{Gavin2006}, Gene Ontology \cite{Ashburner2000,GeneOntology2021}, STRING \cite{Szklarczyk2019}, BioGRID \cite{Stark2006}, and the EBI Complex Portal \cite{Meldal2022}, as described in the Methods. Raw third-party datasets should be obtained from their original providers subject to their respective licences.

\bmhead{Acknowledgements}
We thank the STRING, Gene Ontology, and Complex Portal consortia for public data resources. We also thank the Centre of Research and Funding (CoRFu) at SRH University Heidelberg for funding-advisory and proposal-support guidance.

\bmhead{Author contributions}
S.S., R.S., M.J., and Y.F. contributed to conceptualisation and methodology. S.S. and Y.F. implemented the computational workflow and performed the experiments. R.S. contributed to validation, interpretation of results, and manuscript review. M.J. supervised the study and contributed to methodological design, interpretation of results, manuscript revision, and project coordination. All authors contributed to writing, reviewed the manuscript, and approved the final version.

\bmhead{Funding}
No specific funding was received for this study.

\bmhead{Declarations}

\textbf{Conflict of interest} The authors declare that they have no known competing financial interests or personal relationships that could have appeared to influence the work reported in this paper.

\textbf{Generative AI declaration} During manuscript preparation, AI-assisted tools may have been used only for language editing, formatting support, and editorial refinement. The authors reviewed and edited all content and take full responsibility for the scientific accuracy, integrity, and final content of the manuscript.


\end{document}